\documentclass[12pt]{iopart}

\usepackage{iopams}

\usepackage{tikz}
\usepackage{ifthen}

\newcommand{\xw}[3]{%
  \ifthenelse{\equal{#3}{888888}}{}{
  \ifthenelse{\equal{#3}{666666}}{{\node at (#1,#2) {$\circ$};}}{
  \ifthenelse{\equal{#3}{999999}}{{\node at (#1,#2) {$\bullet$};}}
  {\node at (#1,#2) {$#3$};}}}}

\begin{document}

\title[Algebraic entropy computations for lattice equations]{Algebraic
  entropy computations for lattice equations: why initial value
  problems do matter\\ }

\author{J. Hietarinta$^1$, T. Mase$^2$ and R. Willox$^2$}
\address{$^1$Department of Physics and Astronomy, University of Turku,
  FIN-20014 Turku, Finland}

\address{$^2$Graduate School of Mathematical Sciences, the University
  of Tokyo, 3-8-1 Komaba, Meguro-ku, Tokyo 153-8914, Japan}

\begin{abstract}
In this letter we show that the results of degree growth (algebraic
entropy) calculations for lattice equations strongly depend on the
initial value problem that one chooses. We consider two
  problematic types of initial value configurations, one with problems in
  the past light-cone, the other one causing interference in the
  future light-cone, and apply them to Hirota's discrete KdV equation
  and to the discrete Liouville equation.  Both of these initial
value problems lead to exponential degree growth for Hirota's dKdV,
the quintessential integrable lattice equation.  For the discrete
Liouville equation, though it is linearizable, one of the initial
value problems yields exponential degree growth whereas the other is
shown to yield non-polynomial (though still sub-exponential)
growth. These results are in contrast to the common belief that
discrete integrable equations must have polynomial growth and that
linearizable equations necessarily have linear degree growth,
regardless of the initial value problem one imposes.  Finally,
as a possible remedy for one of the observed anomalies,
we also propose basing integrability tests that use growth criteria
on the degree growth of a single initial value instead of all the
initial values.

\end{abstract}


\section{Introduction}
By now it is common knowledge that discrete integrable systems possess
some beautiful underlying mathematical structures. For example, in the case of
integrable bi-rational mappings, insights from algebraic
geometry have led to the development of rigorous criteria (and tests)
for integrability
\cite{Veselov92,FalquiViallet93,Takenawa01,Maseetal19}, and to a
widely accepted definition of an integrable mapping: it must have zero
algebraic entropy \cite{BellViall99}. The algebraic entropy for a
rational mapping is defined as
\[
E=\lim_{n\to\infty}\frac{\log(d_n)}n,
\]
where $d_n$ is the degree of the $n$th iterate of the mapping,
  and can be thought of as measuring the complexity of the mapping, in
  the sense of Arnold \cite{arnold}.  Note that $E\geq0$ and that it
is non-zero only if the degree growth is exponential.

The situation for lattice equations, however, is very different and a
definition of integrability in that context remains elusive. (As a
general reference see \cite{HJN}.)  Attempts have been made to
introduce to the lattice setting certain integrability criteria that
proved useful in the context of mappings, such as algebraic entropy
\cite{TGR01,Viallet2006,RobertsTran19}, but in this letter we wish to
warn against any such attempt that does not take into account the
crucial role that initial value problems play in the lattice setting.

\subsection{The lattice setting}
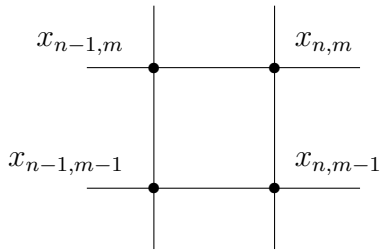
\begin{figure}[h]
\begin{center}
\setlength{\unitlength}{0.00035in}
\begin{picture}(5482,4800)(1000,0)
\put(3275,2708){\circle*{150}}
\put(1525,2900){\makebox(0,0)[lb]{$x_{n-1,m}$}}
\put(5075,2708){\circle*{150}}
\put(5375,2900){\makebox(0,0)[lb]{$x_{n,m}$}}
\put(3275,908){\circle*{150}}
\put(1100,1108){\makebox(0,0)[lb]{$x_{n-1,m-1}$}}
\put(5075,908){\circle*{150}}
\put(5375,1108){\makebox(0,0)[lb]{$x_{n,m-1}$}}
\put(2275,2708){\line(1,0){4075}}
\put(5075,3633){\line(0,-1){3633}}
\put(2275,908){\line(1,0){4075}}
\put(3275,3633){\line(0,-1){3633}}
\end{picture}
\end{center}
\caption{The elementary square in the $\mathbb Z^2$ lattice. The
  independent variable $n$ grows to the right, $m$ grows
  upward. \label{F:1}}
\end{figure}

In this letter we only consider lattice equations, i.e. partial
difference equations, defined on the basic elementary square or
quadrilateral of the $\mathbb Z^2$ lattice, see Figure \ref{F:1}. Such
equations are called {\it quad equations}. The four corner values of
the $2\times 2$ stencil are related by a multi-linear equation
\begin{equation}\label{eq:Q}
Q(x_{n-1,m-1},x_{n,m-1},x_{n,m-1},x_{n,m})=0,
\end{equation}
which may also contain various parameters.

\begin{figure}
\centering
\begin{tikzpicture}[scale=0.8]
\foreach \x in {0,1,2}{%
      \foreach \y in {0,1,2}{%
        \node[draw,circle,inner sep=1.5pt] at (\x,\y) {};}}
\foreach \x in {0,1,2} \node[draw,circle,inner sep=1.5pt,fill] at (\x,-1) {};
\foreach \y in {-1,0,1,2} \node[draw,circle,inner sep=1.5pt,fill] at (-1,\y) {};
\foreach \y in {-2,-1,...,2}%
 \draw[thin] (-2.5,\y) -- (2.5,\y);
\foreach \x in {-2,-1,...,2}%
 \draw[thin] (\x,-2.5) -- (\x,2.5);
\draw[very thick] (-1,-1) -- (-1,2.5);
\draw[very thick] (-1,-1) -- (2.5,-1);
\node at (-0.5,-3) {a)};
\end{tikzpicture}\hspace{1.5cm}\begin{tikzpicture}[scale=0.8]
\foreach \x in {-2,-1,...,2}
        \node[draw,circle,inner sep=1.5pt,fill] at (\x, - \x) {};
\foreach \x in {-2,-1,...,1}
        \node[draw,circle,inner sep=1.5pt,fill] at (\x, - \x-1) {};
\foreach \z in {-1,0,1,2}{%
\foreach \x in {\z,...,2}
        \node[draw,circle,inner sep=1.5pt] at (\x,-\x+\z+2) {};}
\foreach \x in {-2,-1,...,1}
       \draw[very thick] (\x,-\x) --  (\x,-\x-1); 
\foreach \x in {-2,-1,...,1}
       \draw[very thick] (\x,-\x-1) --  (\x+1,-\x-1); 

\draw[very thick] (-2,2) -- (-2.5,2);
\draw[very thick] (2,-2) -- (2,-2.5);

\foreach \y in {-2,-1,...,2}%
 \draw[thin] (-2.5,\y) -- (2.5,\y);
\foreach \x in {-2,-1,...,2}%
 \draw[thin] (\x,-2.5) -- (\x,2.5);

\node at (-0.5,-3) {b)};
\end{tikzpicture}
\caption{a) The Cartesian lattice with given corner-type initial values (at black
  discs) from which one can then compute the values at open discs in the
  upper right quadrant. b) The same with a staircase of initial
  values.\label{F:2}}
\end{figure}
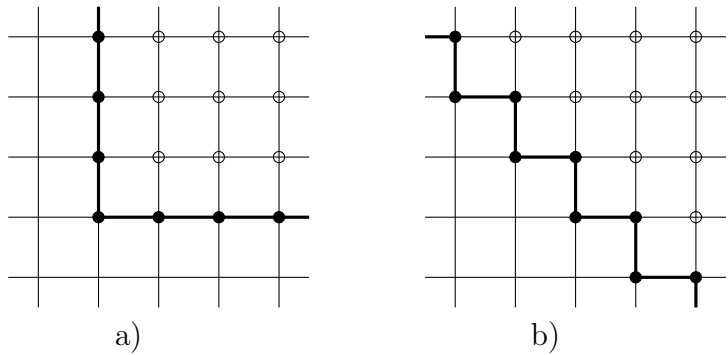

Any multi-linear quad equation allows propagation or evolution once
suitable initial data is given, for example on a staircase or on a
corner, as in Figure \ref{F:2}, but here we shall also consider some
more exotic initial value problems. Note however that all initial value problems we shall consider are {\em well-posed} in the sense of \cite{AdlerVeselov04}.

In this letter we illustrate our results using two quad equations:
\begin{itemize}
\item {\bf Liouville equation:} 
\begin{equation}\label{eq:Liouv}
x_{n,m} x_{n-1,m-1}-x_{n,m-1} x_{n-1,m}=1,
\end{equation}
which linearizes to (non-autonomous) second order ordinary difference equations, in the $n$ as well as in the $m$ direction.
\item {\bf Hirota's discrete KdV equation}
\begin{equation}\label{eq:HKdV}
x_{n,m}-x_{n-1,m-1}=\frac1{x_{n,m-1}}-\frac{\lambda}{x_{n-1,m}},
\end{equation}
which is integrable for $\lambda=1$ but non-integrable for any other non-zero value of $\lambda$.
\end{itemize}

\subsection{Algebraic entropy}
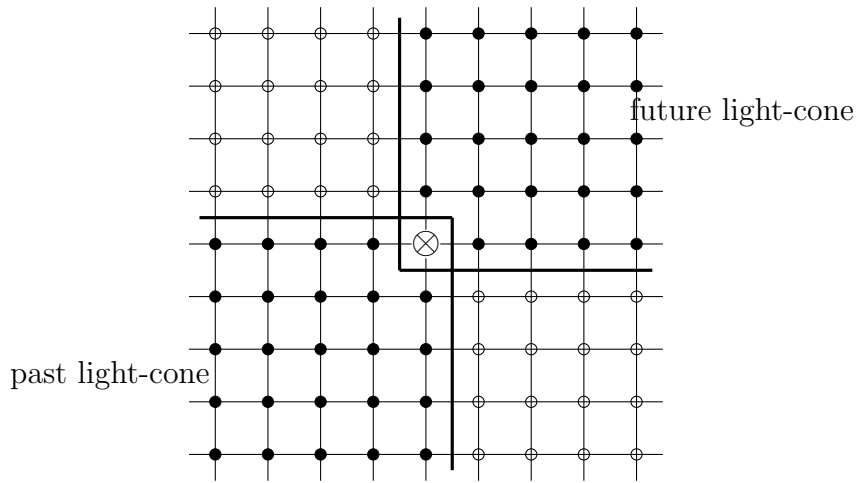
\begin{figure}
\centering
\begin{tikzpicture}[scale=0.7]
\foreach \x in {-5,-4,...,-2}{%
      \foreach \y in {0,1,...,3}{%
        \node[draw,circle,inner sep=1.5pt] at (\x,\y) {};}}
\foreach \x in {0,1,...,3}{%
      \foreach \y in {-5,-4,...,-2}{%
        \node[draw,circle,inner sep=1.5pt] at (\x,\y) {};}}
\foreach \x in {-1,0,1,...,3}{%
      \foreach \y in {-1,0,1,...,3}{%
 \node[draw,circle,inner sep=1.5pt,fill] at (\x,\y) {};}}
\foreach \x in {-5,-4,...,-1}{%
      \foreach \y in {-1,-2,...,-5}{%
 \node[draw,circle,inner sep=1.5pt,fill] at (\x,\y) {};}}

\foreach \y in {-5,-4,...,3}%
 \draw[thin] (-5.5,\y) -- (3.5,\y);
\foreach \x in {-5,-4,...,3}%
 \draw[thin] (\x,-5.5) -- (\x,3.5);
\draw[very thick] (-1.5,-1.5) -- (-1.5,3.3);
\draw[very thick] (-1.5,-1.5) -- (3.3,-1.5);
\draw[very thick] (-0.5,-0.5) -- (-0.5,-5.3);
\draw[very thick] (-0.5,-0.5) -- (-5.3,-0.5);
\node[draw,circle,color=white,inner sep=3.7pt,fill] at (-1,-1) {};
\node[draw, opacity=0,text opacity=1] at (-1,-1) {\large$\otimes$};
\node at (-7,-3.5) {past light-cone};
\node at (5,1.5) {future light-cone};
\end{tikzpicture}
\caption{Past and future light-cones, for equations (\ref{eq:Liouv})
  and (\ref{eq:HKdV}), for the point $\otimes$ in the middle.  The
  past light-cone contains all points that can influence the value at
  $\otimes$, while the future light-cone contains all points that can
  be influenced by the value at $\otimes$.\label{F:3}}
\end{figure}

As already mentioned, one property that is believed to
be strongly associated with integrability is {\it zero Algebraic
  Entropy}. For lattice equations (\ref{eq:Liouv}) or (\ref{eq:HKdV}),
given initial values such as in Figure \ref{F:2}, $x_{n,m}$ at any of
the open discs will be a rational function of the initial values given
at the black discs in the past light-cone of that open disc,
c.f. Figure \ref{F:3}.  The degrees of the numerator and denominator
of $x_{n,m}$ will grow as $n$ and $m$ grow, but if the system is
integrable there will be cancellations and the growth slows down
\cite{Veselov92,FalquiViallet93,HV98,TGR01}.

In \cite{TGR01} it was shown that for initial values as in Figure 2,
Hirota's discrete KdV equation (\ref{eq:HKdV}) exhibits quadratic
degree growth whereas the discrete Liouville equation (\ref{eq:Liouv})
has linear degree growth (see also section 2.1). In fact, besides
exponential growth (i.e.  non-zero algebraic entropy), the degree
growths that have been reported in the literature for lattice
equations \cite{Viallet2006,RobertsTran19} coincide with those
observed for second order bi-rational mappings: bounded growth, linear
growth or quadratic degree growth. Hence the `rule of thumb' that is
in common use today
(cf. \cite{FalquiViallet93,BellViall99,RGLO2000,TGR01,HV07,Gub19})
\begin{itemize}
\item if the degree growth is linear in $n,m$ the equation is
  linearizable.
\item if the degree growth is {\it polynomial} the equation is
  integrable.
\item if the degree growth is exponential the equation is
  not integrable.
\end{itemize}
It is important to stress that although the above statements can be
considered to be rigorous in the context of second order bi-rational
mappings \cite{dillerfavre}, when it comes to lattice equations these
should be thought of as mere experimental observations.  

In this
letter we will discuss how the form of the initial value boundary may
influence degree growth computations and distort the algebraic entropy
results. We concentrate on two particular effects that have their
origin in:
\begin{enumerate}
\item The number of initial values in the past light-cone. Usually this number grows linearly, but it may in fact grow faster.
\item The number of initial values in the future light-cone. Usually
  there are none but for some allowed (well-posed) initial value
  problems this number can be non-zero.
\end{enumerate}
In this letter we will only discuss what can happen for the Liouville
equation (\ref{eq:Liouv}) and for Hirota's discrete KdV
(\ref{eq:HKdV}) equation, both defined on a $2\times2$ stencil, and we
shall report work on other kinds of stencils elsewhere.

\section{Growth results for various initial value problems}
\subsection{Corner initial values}
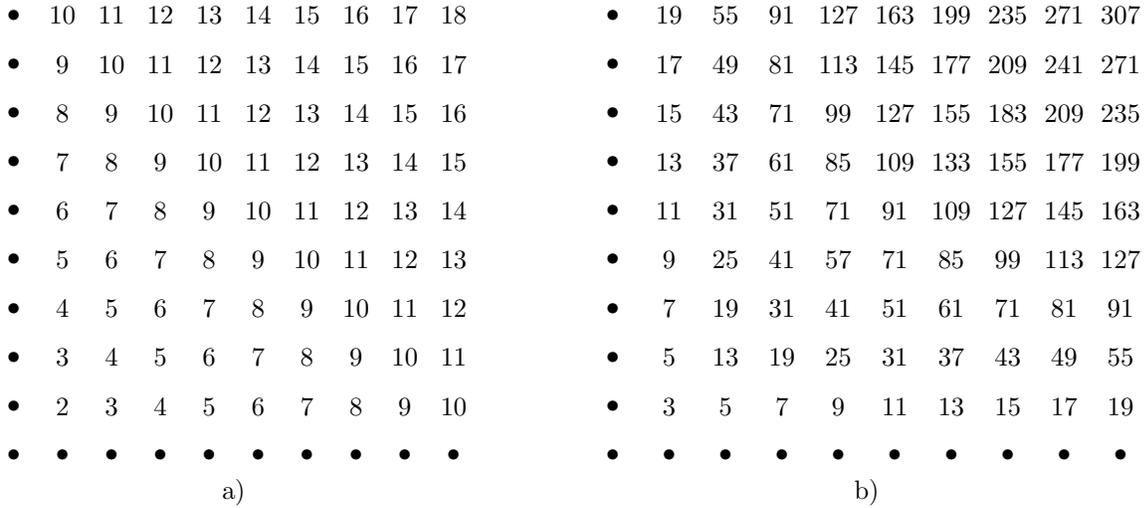
\begin{figure}[h]
  \centering
  {\footnotesize
    \begin{tikzpicture}[scale=0.65]
 \input 1corner10tikzliouv
\node at (4.5,-0.75) {a)};
    \end{tikzpicture}\hspace{1.5cm}
    \begin{tikzpicture}[xscale=0.75,yscale=0.65]
  \input 1corner10tikz
\node at (4.5,-0.75) {b)};
\end{tikzpicture}
  }
  \caption{a) For the Liouville equation (\ref{eq:Liouv}) the degree growth
    is linear, as expected: $d_{n,m}=n+m$. b) Hirota's KdV equation
    (\ref{eq:HKdV}) for $\lambda=1$ is integrable and the degree rule
    is $d_{n,m}=4nm-2\max(n,m)+1$ (see also \cite{TGR01})\label{F:LH1}.}
\end{figure}

\begin{figure}
  \centering
  {\footnotesize
\begin{tikzpicture}[xscale=1.1,yscale=0.6]
  \input  2corner10tikz
\end{tikzpicture}}
    \caption{Hirota's KdV with $\lambda=2$ (non-integrable). On the
      diagonal the growth is exponential, approximately
      $0.85\cdot3.7^n$. It can be shown however (see section 2.3) that the asymptotic degree growth on the diagonal is actually $\propto4^n$.
  \label{F:Hn1}}
\end{figure}
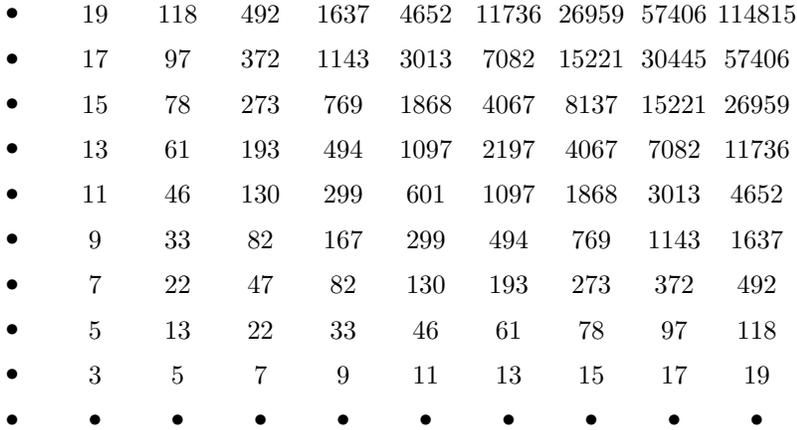

Let us first consider conventional corner initial values as in Figure
\ref{F:2}a.  Figure \ref{F:LH1} shows the degrees of the numerator in
$x_{n,m}$ for the Liouville equation (\ref{eq:Liouv}), which are given
by $d_{n,m}=n+m$, and for Hirota's KdV equation (\ref{eq:HKdV}) with $\lambda=1$: 
$d_{n,m}=4nm-2\max(n,m)+1$ (cf. \cite{TGR01}). As before,
black discs depict initial values.
 Figure \ref{F:Hn1} gives the
degrees for a non-integrable version ($\lambda=2$) of Hirota's KdV
equation and this time the growth is exponential. The first difference
between the integrable and non-integrable versions occurs at lattice
points $(2,3)$ and $(3,2)$ where the integrable version has a
numerator of degree 19 while the non-integrable one has degree 22. One
can easily compute the corresponding rational functions $x_{n,m}$. It
turns out that for any $\lambda$ the denominator of $x_{3,2}$ has the
degree 3 factor
\[
x_{1,0}\ x_{0,1}\ x_{0,0} - x_{1,0}\ \lambda + x_{0,1}.
\]
If we now insist that this should also be a factor of the numerator
(in order to have cancellations) one finds that this is only
possible if $\lambda=1$ or $0$.

\subsection{Problems originating in the past light-cone}
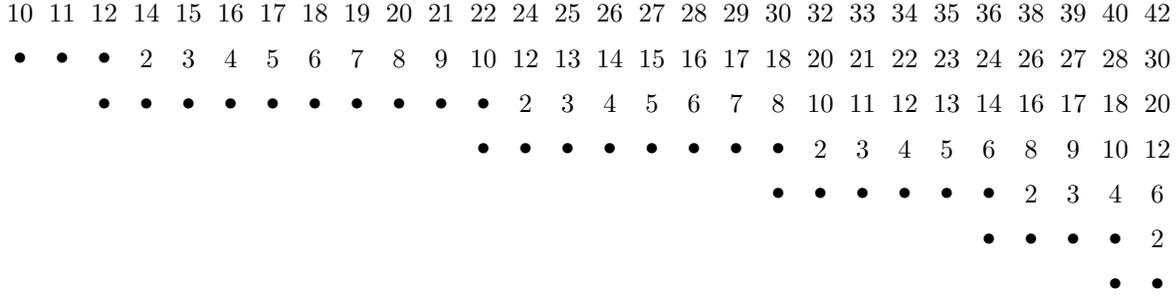
\begin{figure}
    \centering
    {\footnotesize
  \begin{tikzpicture}[yscale=0.6,xscale=0.56]
  \input paraliouv7tikz
\end{tikzpicture}
    }
    \caption{Liouville equation for a left-leaning parabolic
      boundary. Bullets depict the initial values. The degrees in the
      rightmost column grow as $m(m+1)$.
  \label{F:Lpast1}}
\end{figure}

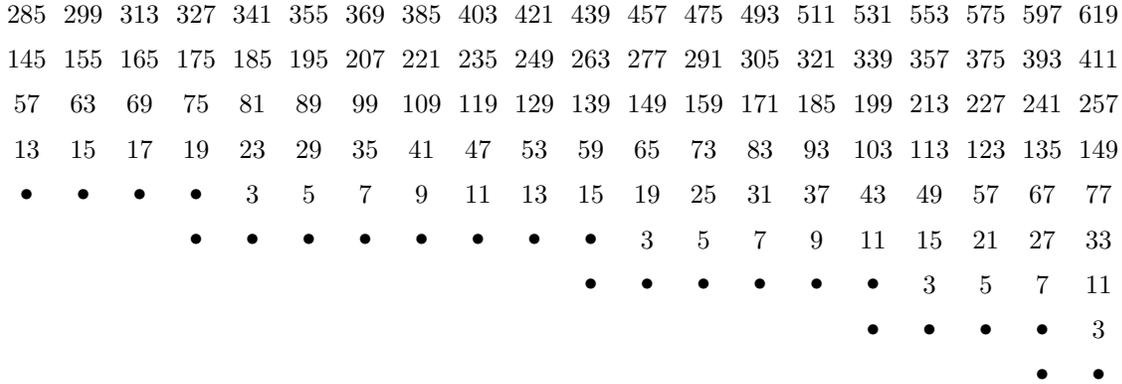
\begin{figure}
    \centering
{\footnotesize
  \begin{tikzpicture}[xscale=0.75,yscale=0.6]
  \input parahkdv11tikz
  \end{tikzpicture}
  }
\caption{Hirota's KdV with parabolic boundary to the left. The growth
  in the rightmost column is now cubic,
  $\frac16(8m^3-9m^2+25m-9)\pm\frac12$.
\label{F:Hpast1}}
\end{figure}

For a quad equation and with evolution to the NE direction, the
initial values $f_{\alpha,\beta}$ that appear in $x_{n,m}$ are those
in its past light-cone, except that of the $f_{\alpha,m}$ the only initial
value that is included is the first to the left.  Possible problems
with the past light cone are illustrated in Figures \ref{F:Lpast1} and
\ref{F:Hpast1}.
Let us first take a closer look at the computations in the case of the
Liouville equation. In Figure 6 we use numbering in which the
rightmost column is at $n=0$. For the first two points in that column
we find
\begin{eqnarray}
  x_{0,1}&=&\frac{f_{0,0}\ f_{-1,1}+\lambda}{f_{-1,0}},\\
  x_{0,2}&=&\frac{f_{0,0}\ {f_{-1,1}}^2\ f_{-2,1}\ f_{-3,1}\ f_{-4,2} + l.o.}
  {f_{-1,1}\ f_{-1,0}\ f_{-2,1}\ f_{-3,1}\ f_{-4,1}} .
\end{eqnarray}
Thus $x_{0,2}$ depends on all initial values of type $f_{n,0}$ and
$f_{n,1}$ but only on $f_{-4,2}$ in the $m=2$ row, i.e. it depends on
7 initial values and the highest degree term depends on exactly 6 of
them.

 Usually the number of initial values in the past light-cone grows
 linearly with $n$ or $m$; this is what happens for the corner and
 staircase initial configurations of Figure \ref{F:2}.  However, if
 the number of initial values grows quadratically with $m$ (as in the
 case of the boundaries in Figures \ref{F:Lpast1} and \ref{F:Hpast1}
 where it grows as $m^2+m+1$) then complexity grows one unit faster:
 the degree growth in the $m$ direction is quadratic for Liouville and
 cubic for Hirota's KdV, as can be seen from Figure \ref{F:Hpast1}.
 But the complexity can be made to grow exponentially for Hirota's KdV
 or even for Liouville, if the initial values for the left-hand
 boundary are located for example at $-2^m\le n\le -2^{m-1}$.
 
The effects induced by a faster-than-linear growth of the number of
initial values in the past light-cone, illustrated above, are fairly
interesting by themselves. However, if we wish to use degree growth as
an objective criterion and, most importantly, as a tool for
identifying integrable lattice equations then these effects are best
eliminated.

\begin{figure}[h]
    \centering
      {\footnotesize
\begin{tikzpicture}[scale=0.55]
  \input 1ind10tikzliouv
\end{tikzpicture}
      }
      \caption{Degrees w.r.t an individual initial value: Liouville
        equation. With the linear growth of the boundary eliminated
        the degree growth is bounded.\label{F:indL}}
  \end{figure}
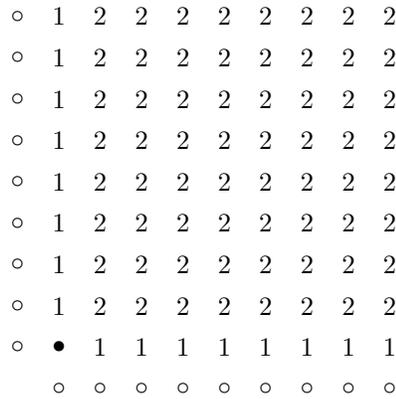

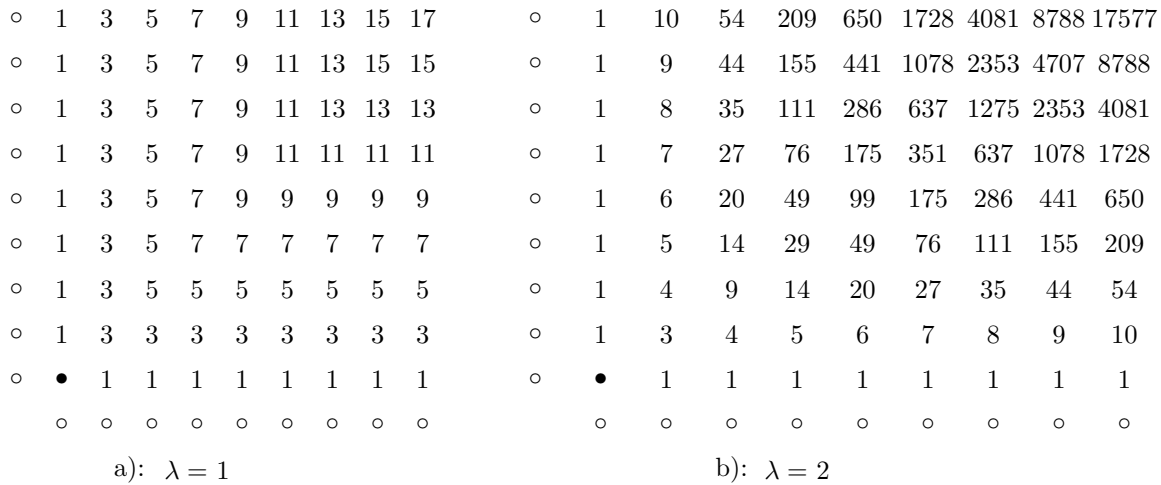
\begin{figure}[b]
    \centering
 {\footnotesize
\begin{tikzpicture}[yscale=0.6,xscale=0.6]
  \input 1ind10tikz
  \node at (2.5,-1) {a):};
\end{tikzpicture}\hspace{1.0cm}\begin{tikzpicture}[yscale=0.6,xscale=0.87]
  \input 2ind10tikz
    \node at (3,-1) {b):};
\end{tikzpicture}
  }
\caption{Degrees w.r.t an individual initial value: Hirota's KdV for
  $\lambda=1$ (integrable) and $\lambda=2$ (non-integrable). When the
  linear growth of the boundary is eliminated the degrees in the
  integrable case grow linearly, while for the non-integrable cases
  the degrees still grow exponentially. }\label{F:indH1}
\end{figure}

One solution would be to consider the degree growth with respect to
{\it only a single initial value.} (In practical computations all
other initial values can then be numerical.) Such an approach is
illustrated in Figures \ref{F:indL} and \ref{F:indH1}.
 In each of
these figures the missing corner point is taken to be at $(0,0)$ and
the sole initial value at $(1,1)$. The initial values $f_{1,m}\,
(m>1)$ and $f_{n,1}\, (n>1)$ are numerical. For the Liouville equation
the degrees are then bounded ($d_{n,m}=\min(n,m,2)$) whereas for the
usual, integrable, version of Hirota's KdV the degrees in Figure
\ref{F:indH1}a) are given by $d_{n,m}=2\min(n,m)-1$, i.e. they now
exhibit linear growth.
 For the non-integrable version of Hirota's KdV
equation of Figure \ref{F:indH1}b) the growth is still exponential.  It
is interesting to observe that the degrees in Figures \ref{F:indH1}b)
and \ref{F:Hn1} satisfy very similar recursion relations:
$$d_{n,m}= d_{n-1,m} + d_{n,m-1} ~~(n\neq m)\quad{\rm and}\quad d_{n,n} = d_{n-1,n} + d_{n,n-1}+1,$$
for Figure \ref{F:indH1}b) and
$$d_{n,m}= d_{n-1,m} + d_{n,m-1} +2 ~~(n\neq m)\quad{\rm and}\quad
d_{n,n} = d_{n-1,n} + d_{n,n-1}+3,$$ for Figure \ref{F:Hn1}. In fact,
the number sequence on the diagonal in Figure \ref{F:indH1}b) is
known, it is sequence M2811 or A006134 in \cite{sloane-p:encyclopedia}
and it can be shown to grow asymptotically as $4^n$.  As the only
difference between the settings in Figure \ref{F:indH1}b) and Figure
\ref{F:Hn1} is that in the latter the number of initial values grows
linearly with $n$ and $m$, whereas in Figure \ref{F:indH1}b) there is
only a single initial value, it follows that asymptotically the
degrees on the diagonal in Figure \ref{F:Hn1} should grow at least as
fast as $4^n$. It can be shown that this is indeed the exact
asymptotic growth rate.

In summary, if we only measure the degree growth with respect to a
single specific initial value (as opposed to {\em all} initial values
as one does in standard algebraic entropy calculations) any problems
due to the past light-cone are eliminated. One may say
  that the effects from the past light-cone are ``additive'' and by
  choosing only one initial value as independent variable we can take
  care of this artefact.  An added bonus of this approach is that, in
practice, instead of symbols one can assign numerical values to all
other initial values, which dramatically reduces the calculation time.
Obviously, finding even just a single set of such values for which the
degree growth becomes exponential is sufficient to show that the
equation exhibits exponential degree growth in the usual sense. But if
the aim is to establish integrability by showing that exponential
growth is absent from the system, then of course one has to perform
the calculations with greater care and verify that the result does not
depend on the specifics of the numerical initial values or, indeed, on
the choice of position for the initial value for which the degree is
calculated.

\subsection{Problems originating in the future light-cone}
If the boundary on which the initial values are given intersects the
future light-cone, the effect on the degree growth can be even more
dramatic than in the previous case. With such a boundary there can be (from the viewpoint of
integrability tests) potentially catastrophic interference when new
independent variables---i.e. initial values---prevent
cancellations. Such interference will result in much faster degree
growth than one would find for ordinary corner or staircase-type
initial value problems.

\begin{figure}[h]
  \centering
  {\footnotesize
\begin{tikzpicture}[xscale=1,yscale=0.55]
  \input h1kdv1212tikz
\end{tikzpicture}
}
\caption{The integrable Hirota dKdV with 45 degree corner
  configuration. Along the horizontal axis the growth is linear after
  the first 2 lattice points, but along the diagonal the growth is
  exponential $d_{n,n}=3(2^n-1)$ and next to it $9(2^{n-1}-1)-2n$.
  \label{F:H45}}
\end{figure}
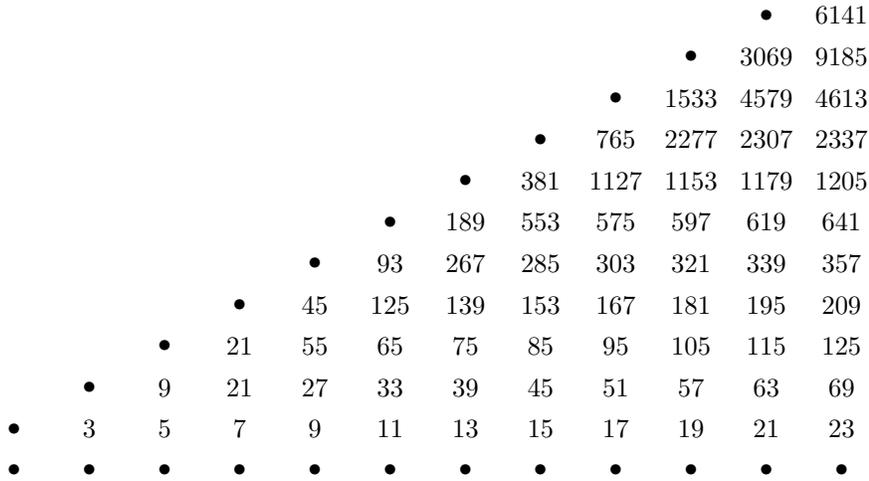

One such case is illustrated in Figure \ref{F:H45} for Hirota's
discrete KdV equation.
  Instead of a vertical boundary as one would have in a corner-type
  initial value problem, we now have a $45^o$ forward leaning
  boundary. Note that the upper boundary is not a staircase and
  therefore evolution is well defined, i.e. the initial value problem
  is well-posed. (A staircase would allow evolution to the SE
  direction, conflicting eventually with evolution starting from the
  bottom boundary.)  We observe that on the diagonal the degree growth
  is exponential, $d_{n,n}=3(2^n-1)$, and that one step to the right
  from the diagonal the degrees are given by
  $d_{n,n-1}=9(2^{n-1}-1)-2n$.  After these first two steps the degree
  growth for a given $m$ as a function of $n$ is the same as in Figure
  \ref{F:LH1} b), namely $(4m-2)n$. We note that for the Liouville
  equation the degrees in the $45^o$ corner region are in fact the
  same as in the $90^o$ corner of Figure \ref{F:LH1}a).
 
\begin{figure}[t]
  \centering
  {\footnotesize
\begin{tikzpicture}[xscale=1,yscale=0.55]
  \input 1strfor112tikz1
\end{tikzpicture}
}
\caption{The integrable Hirota dKdV with 45 degree corner
  configuration and a single variable . Along the diagonal the growth is
  exponential $d_{m,m}=2^{m-1}$ and for $n>m$ $d_{n,m}=3\ 2^{m-1}-2$.
    \label{F:H112}}
\end{figure}
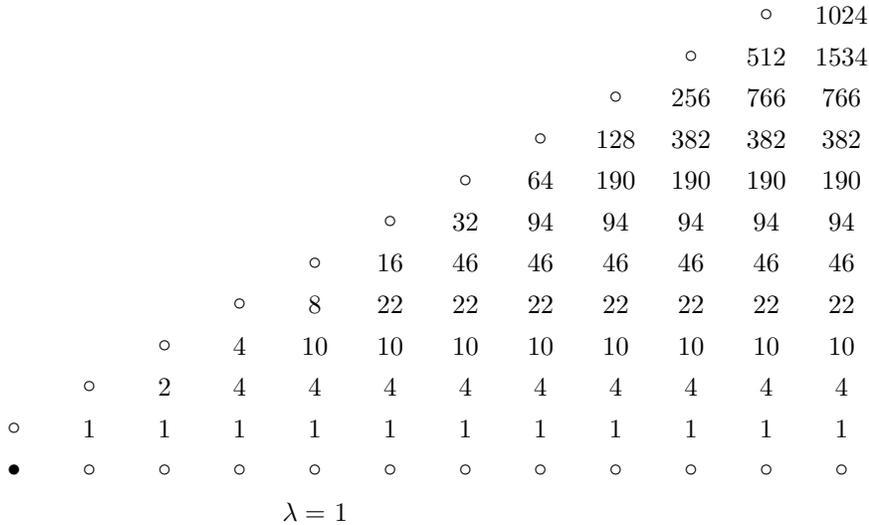

Above we have found that the {\em past} light-cone artefacts
can be eliminated by calculating degrees with respect to just one independent variable among the
  initial values.  It should be clear however that the same approach will not resolve problems that have to do with the future light-cone. For example, if we calculate the degree growth with respect to a single intial value for the initial
  value problem of Figure \ref{F:H45} we get the result in Figure
  \ref{F:H112}.
Observe that the growth rate on the diagonal $m=n$ is $2^m$ for both
cases. Though degree growth calculations with respect to a single
initial value do not remove the problem, they do yield results that
display it in a distilled form.

As mentioned before, when computing the degree growth with respect to
a single initial variable all the other initial values can be taken as
(generic) numerical values, and the degrees in Figure 11 are in fact
obtained in this way.  One could of course wonder why in such a
calculation where all but a single initial value take numerical
values, the growth on the diagonal in Figure \ref{F:H112} is so much
faster than that on the diagonal for the corner initial value problem
of Figure \ref{F:indH1}. But it should be noted that the values
derived from the corner configuration are intricately related,
allowing subtle cancellations, which is the reason for slow growth. If
we replace these variables by random numbers the relations are
eliminated and cancellations are no longer possible. (One can say that
the future light-cone problems are ``multiplicative'' in nature.) Vice
versa, we can note that the initial conditions on the vertical
boundary in Figure \ref{F:indH1} needed to reproduce these exact
numerical values on the line $ m=n+1$ are in fact increasingly complex
rational expressions in the (sole) symbolic initial value.

\begin{figure}[b]
  \centering
    {\footnotesize
\begin{tikzpicture}[scale=0.65]
  \input slopeliouv318tikz
  \draw[thin,opacity=0.125,fill=black] (1,0.5) -- (16.5,16) --
  (16,16.5) -- (0.5,1) ;
\end{tikzpicture}
    }
    \caption{Discrete Liouville equation in a step 3 wedge.\label{F:Lwedge1}}
\end{figure}
\begin{figure}
  \includegraphics[scale=0.4]{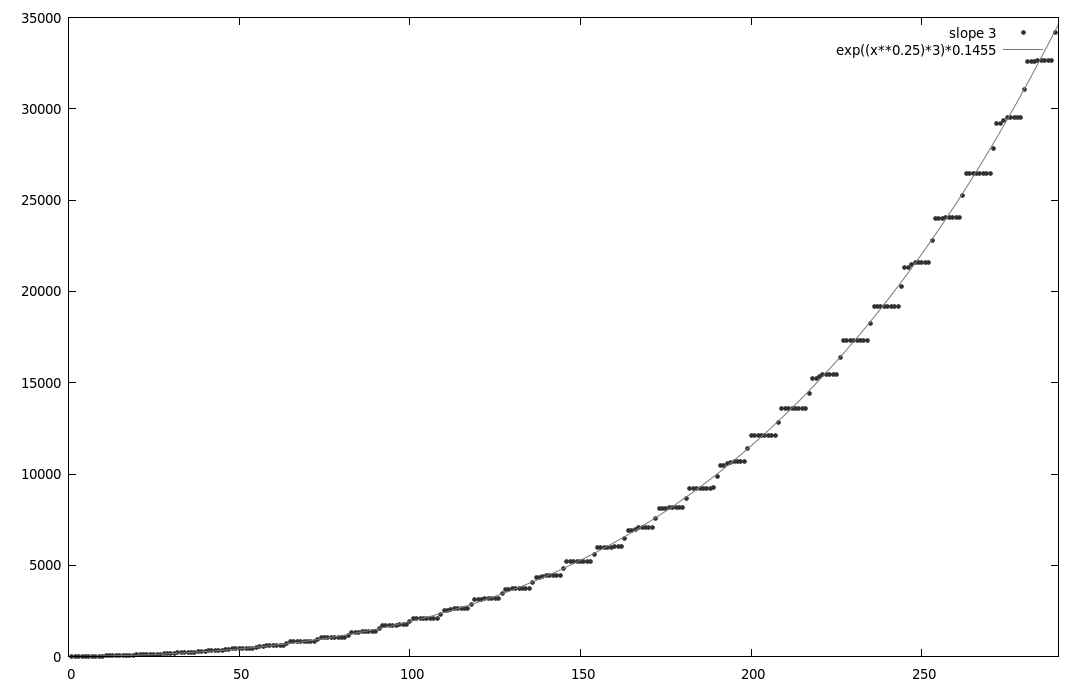}
  \caption{Same situation as in Figure \ref{F:Lwedge1} but for
    extended range.  The vertical axis gives the degree of the
    numerator and the horizontal axis the order of iteration $n$.  The
    approximate fit is with $d_{n,n}=0.1455\cdot\exp[3n^{\frac14}]$.
       \label{F:Lwedge2}}
\end{figure}

We have also studied some other initial boundaries that can be
problematic. One possibility is to have a wedge type boundary as in
Figure \ref{F:Lwedge1} for the discrete Liouville equation. From the
degrees on the shaded diagonal it is, at first, difficult to determine
the rule for the degree growth but after computing more steps the type
of growth on the diagonal becomes clearer, see Figure
\ref{F:Lwedge2}. There are regions of slow growth followed by jumps
and we get a fairly good fit for the degree growth with the curve
$0.1455\cdot\exp[3x^{\frac14}]$, but similar fits with slightly
different exponents are also possible. In all cases the exponent was
found to be close to $x^{\frac14}$ and thus asymptotically the growth
is sub-exponential but faster than any polynomial: $a n^\alpha <
d_{n,n} < b e^{\beta n}$ for all $\alpha,\beta >0$ (and suitable
constants $a(\alpha),b(\beta)$), and the algebraic entropy as
conventionally defined is still zero.

\section{Summary}
In this letter we have discussed how the choice of initial value
problem can affect the degree growth in a lattice
equation. Conventional initial value problems use corner or staircase
boundaries, but one can have a well defined (well-posed) evolution
starting with other kinds of initial value boundaries.

One of the problems we have identified concerns the number of initial
values in a point's past light-cone, which influences the subsequent
degrees. In particular, for a boundary that recedes exponentially, we
found that one can obtain exponential degree growth even for
linearizable equations.  In order to avoid any such ambiguous growth
due to past light-cone effects, we propose to study the degree growth
with respect to one individual initial value instead of all initial
values. If this growth is polynomial, then its degree is one less than
that for the conventional corner and staircase initial
values. Moreover, in such a calculation the remaining initial values
may be taken to be purely numerical, but in that case sufficient care
must be taken to ensure that the choice of initial values does not
influence the observed degree growth (e.g., due to accidental
cancellations).

The situation is even more interesting and problematic 
in cases where the initial value boundaries intersect the future
light-cone. We have shown that if, for Hirota's discrete KdV equation,
the boundary of the corner configuration is tilted forward to a $45^o$
angle, one can observe exponential degree growth even though the
lattice equation is supposedly integrable.  Another interesting
finding is the behaviour in the wedge between sloping boundaries as in
Figure \ref{F:Lwedge1}. For the Liouville equation we find growth that
is faster than polynomial but still sub-exponential. As far as the
authors know this is the first time this type of degree growth has
been observed in lattice equations.

As we have shown, when testing the degree growth for a lattice
equation, it is imperative that one avoid initial value problems in
which initial values appear in the future light-cones of the lattice
points that one wishes to calculate. (It is not immediately clear
however whether this simple safeguard is actually sufficient or
whether there exist other boundary-induced effects that should be
taken into account.)

In this letter we have presented results for two quad equations, the
discrete Liouville equation and Hirota's discrete KdV. However, we have
obtained similar results also for Hirota's bilinear KdV equation
($2\times 3$ stencil) and for the bilinear Toda lattice equation
(star-shaped, 5 points.)  For the Toda equation we have several
rigorous results for the degrees which will be presented elsewhere.

While we have been discussing integrability from the point of view of
the growth of complexity, there are other viewpoints as well,
e.g. relying on symmetries and Lax pairs. One can then again ask
whether a given initial value problem allows the construction of a
sufficient number of symmetries or of a meaningful Lax pair
\cite{Habi98,Habi01,CCZ,Habi19}. These approaches are for future
consideration.

\ack RW and TM would like to acknowledge support from the Japan
Society for the Promotion of Science (JSPS), through JSPS grants
number 18K03355 and 18K13438, respectively. JH would like to thank
Claude Viallet for advice in degree computations.

\end{document}